\documentclass[a4paper,12pt]{article}

\begin{document}
\topmargin 0pt \oddsidemargin 0mm

\renewcommand{\thefootnote}{\fnsymbol{footnote}}
\begin{titlepage}
\begin{flushright}
\end{flushright}

\vspace{2mm}
\begin{center}
{\Large \bf Emergence of Space and Spacetime Dynamics of
Friedmann-Robertson-Walker Universe}
 \vspace{12mm}

{\large Rong-Gen Cai\footnote{e-mail address: cairg@itp.ac.cn}}

\vspace{5mm} {\em  State Key Laboratory of Theoretical Physics, \\
Institute of Theoretical Physics, Chinese
Academy of Sciences, \\
 P.O. Box 2735, Beijing 100190, China} \\

\end{center}

\vspace{45mm}
\centerline{{\bf{Abstract}}} \vspace{5mm}
 In a recent paper [arXiv:1206.4916] by T. Padmanabhan, it was
 argued that our universe provides an ideal setup to stress the issue that
 cosmic space is emergent as cosmic time progresses and that the
 expansion of the universe is due to the difference between the
 number of degrees of freedom on a holographic surface and the one
 in the emerged bulk. In this note following this proposal we
 obtain the Friedmann equation of a higher dimensional
 Friedmann-Robertson-Walker universe. By properly modifying
 the volume increase and the number of degrees of freedom on the
 holographic surface from the entropy formulas of black hole in the
 Gauss-Bonnet gravity and more general Lovelock gravity, we also
 get corresponding dynamical equations of the universe in those
 gravity theories.
\end{titlepage}

\newpage
\renewcommand{\thefootnote}{\arabic{footnote}}
\setcounter{footnote}{0} \setcounter{page}{2}


It is generally believed that our spacetime is emergent. However, it
is clearly quite difficult to build spacetime structure from some
suitably defined non-geometric variables. On the other hand,
according to the equivalence principle, gravity is just the dynamics
of spacetime. This implies that gravity is also an emergent
phenomenon. Indeed, the idea that gravity appears as an emergent
phenomenon can date back to the proposal made by Sakharov in
1967~\cite{Sakharov}. In this so-called induced gravity, spacetime
background emerges as a mean field approximation of some underlying
microscopic degrees of freedom, similar to hydrodynamics or
continuum  elasticity theory from molecular physics~\cite{Visser}.
Various aspects of the recent research on the relation between
thermodynamics and gravitational dynamics support such a point of
view (for a review, see \cite{Review}).

In the studies of the connection between thermodynamics and
gravitational dynamics, a lot of attention is payed on how the
gravitational field equations appear from the thermodynamical
viewpoint.  In 1995, Jacobson~\cite{Jac} derived Einstein's field
equations by employing the fundamental Clausius relation $\delta Q =
TdS$ together with the equivalence principle. Here the key idea is
to demand that this relation holds for all the local Rindler causal
horizon through each spacetime point, with $\delta Q$ and $T$
interpreted as the energy flux and Unruh temperature seen by an
accelerated observer just inside the horizon. In this way,
Einstein¡¯s equation is nothing but an equation of state of
spacetime. The Clausius relation also arises in the interpretation
of gravitational field equations as an entropy balance law, $\delta
S_{\rm m}=\delta S_{\rm grav}$, across a null surface~\cite{Pad09}.

In a paper by Verlinde~\cite{Verlinde}, the viewpoint of gravity
being not a fundamental interaction has been further advocated.
Gravity is explained as an entropic force caused by changes in the
information associated with the positions of material bodies.  With
the assumption of the entropic force together with the Unruh
temperature~\cite{Unruh}, Verlinde is able to derive the second law
of Newton; with the assumption of the entropic force together with
the holographic principle and the equipartition law of energy he
obtains the Newton's law of gravitation. Similar observations were
also made by Padmanabhan~\cite{Pad10}. He observed that the
equipartition law of energy for the horizon degrees of freedom
combining  with the thermodynamic relation $S = E/2T$, also leads to
the Newton's law of gravity, here $S$ and $T$ are thermodynamic
entropy and temperature associated with the horizon and $E$ is the
active gravitational mass producing the gravitational acceleration
in the spacetime~\cite{Pad04}.

In the setting of a Friedmann-Robertson-Walker (FRW) universe,
applying the Clauius relation to the apparent horizon of the
universe with any spatial curvature, one is able to derive the
Friedmann equations describing the dynamics of the universe, not
only in the Einstein's general relativity, but also in the
Gauss-Bonnet gravity and more general Lovelock
gravity~\cite{CaiKim}. In this derivation, one assumes that there is
a Hawking temperature $T=1/2\pi r_A$ associated with the apparent
horizon~\cite{CCH}, here $r_A$ is the apparent horizon radius. While
in the entropic force picture of gravity, combining the holographic
principle with the equipartition law of energy and the Unruh
temperature, one also can get the Friedmann equations of the FRW
universe~\cite{CCN}, in which the Komar mass plays the role as the
source to produce the gravitational field.

Note that in these studies, one discusses the gravitational field
equations as the equations of emergent phenomenon, assuming the
existence of spacetime manifold. Clearly it is quite difficult to
stress the issue for ``spacetime itself as an emergent structure".
Time is a parameter to describe the evolution of some dynamical
variables.  It is therefore even more difficult conceptually to
think of time as being emergent from some pre-geometric variables,
compared to space. Very recently, Padmanabhan~\cite{Pad12} argued
that our universe provides a setup to stress the issue that {\it the
cosmic space is emergent as the cosmic time progresses}, because the
cosmic time of a geodesic observer plays a special role, to which
the cosmic microwave background radiation is homogeneous and
isotropic. He argued that the expansion of the universe is due to
the difference between the surface degrees of freedom and the bulk
degrees of freedom in a region of emerged space and successfully
derived the dynamical equation of a FRW universe. In this paper we
would like to further elaborate on this issue and generalize to the
Gauss-Bonnet gravity and more general Lovelcok gravity.

Let us start with the Padmanabhan's observation. He notices that for
a pure de Sitter universe with Hubble constant $H$, the holographic
principle can be expressed in terms of the form
\begin{equation}
\label{eq1} N_{\rm sur}=N_{\rm bulk},
\end{equation}
where $N_{\rm sur}$ denotes the number of degrees of freedom on the
spherical surface with Hubble radius $1/H$, $N_{\rm
sur}=4\pi/{L_p^2H^2}$, where $L_p$ is the Planck length, while the
bulk degrees of freedom obey the equipartition law of energy,
$N_{\rm bulk}=|E|/(1/2)T$, in this paper we set $k_B=1$. If one
takes the
 temperature to be the Hawking temperature $T=H/2\pi$ associated with the Hubble horizon
 and $E$ to be the Komar energy $|(\rho+3p)|V$ contained inside
 the Hubble volume $V=4\pi/3H^3$, and considers $p=-\rho$ for the
 de Sitter space, one immediately from (\ref{eq1}) has $H^2=8\pi
 L_p^2 \rho/3$, which shows the consistence between (\ref{eq1}) and
 Einstein's field equations.

 With the equality (\ref{eq1}), one has $|E|= (1/2)N_{\rm sur}T$,
 which is the standard equipartition law. This equipartition law in
 the gravity context was first found for static spacetime
 in \cite{Pad04}. In the entropic force picture, the equipartition
 law is also used to describe the degrees of freedom on the
 holographic screen~\cite{Verlinde}. In \cite{Pad12}, Padmanabhan
 called the equality (\ref{eq1}) the {\it holographic equipartition}
 because it relates the effective degrees of freedom residing in the
 bulk to the degrees of freedom on the boundary surface.

 The validity of the holographic equipartition (\ref{eq1}) for the pure de Sitter
 leads Padmanabhan to further consider our real universe, which is
 asymptotically de Sitter, as shown by a lot of astronomical observations. The main idea
 is to regard that the expansion of the universe, conceptually equivalent to the
 emergence of space, is being derived towards the holographic equipartition, and
 the basic law governing the emergence of space must relate the emergence of space
 to the different between the number of degrees of freedom in the holographic surface
 and the one in the emerged bulk~\cite{Pad12}. He proposed that in an infinitesimal
 interval
 $dt$ of cosmic time, the increase $dV$ of the cosmic volume is given by
 \begin{equation}
 \label{eq2}
 \frac{dV}{dt}=L_p^2 (N_{\rm sur}-N_{\rm bulk}),
 \end{equation}
 Putting the cosmic volume $V=4\pi/3H^3$, the degrees of freedom on the
 holographic boundary $N_{\rm sur}=4\pi /L_p^2H^2$, the temperature $T=H/2\pi$,
 and the degrees of freedom in the bulk $N_{\rm bulk}$ given in terms of
 the Komar energy in the bulk, he arrived at
 \begin{equation}
 \label{eq3}
 \frac{\ddot a}{a}=-\frac{4\pi L_P^2}{3}(\rho+3p).
 \end{equation}
 This is nothing, but the standard dynamical equation of a FRW universe
 filled by perfect fluid with energy density $\rho$ and pressure $p$.
 With the continuity equation, $\dot \rho +3 H(\rho+p)=0$, integrating
 (\ref{eq3}), one can get the Friedmann equation
 \begin{equation}
 \label{eq4}
 H^2 +\frac{k}{a^2}=\frac{8\pi L_p^2}{3}\rho,
 \end{equation}
 where $k$ is an integration constant, which can be interpreted as the spatial
 curvature of the FRW universe.

One can see from (\ref{eq1}) and (\ref{eq2}) the necessary of the
existence of the cosmological constant in order to have the
asymptotic holographic equipartition~\cite{Pad12}. The existence of
the cosmological constant derives our universe towards the state
with the holographic equipartition. Clearly without the cosmological
constant, such a state cannot be reached. In this note we will not
further discuss the implications of (\ref{eq2}) for dark energy,
instead we will pay attention on the dynamical equations of the
universe from (\ref{eq2}).

 We now first generalize the above derivation process of the dynamical equation of the universe to
 the higher $(n+1)$-dimensional case with $n>3$. In this case the number of degrees of freedom on the
 holographic surface is given by~\cite{Verlinde}
 \begin{equation}
 \label{eq5}
 N_{\rm sur}=\alpha A/L_p^{n-1},
 \end{equation}
 where $A= n\Omega_n/H^{n-1}$ and $\alpha=(n-1)/2(n-2)$ with $\Omega_n$ being the
 volume of an $n$-sphere with unit radius. In this case, we make a minor modification
 for the proposal (\ref{eq2}) as
 \begin{equation}
 \label{eq6}
 \alpha \frac{dV}{dt}=L_p^{n-1}(N_{\rm sur}-N_{\rm bulk}),
 \end{equation}
 where the volume $V=\Omega_n/H^n$. In the case of $(n+1)$-dimensions, the bulk Komar
 energy is~\cite{CCN}
 \begin{equation}
 E_{\rm Komar}=\frac{(n-2)\rho+np}{n-2}V,
 \end{equation}
 and then the bulk degrees of freedom read
 \begin{equation}
 \label{eq8}
 N_{\rm bulk}=-\frac{2E_{\rm Komar}}{T},
 \end{equation}
 where we have added a minus sign in front of $E_{\rm Komar}$, in order to have
 $N_{\rm bulk}>0$, which
 makes sense only in the accelerating phase with $(n-2)\rho +np
 <0$~\cite{Pad12}. For normal matters, one needs not the minus sign.
 Substituting (\ref{eq5}) and (\ref{eq8}) into (\ref{eq6}) with temperature $T=H/2\pi$, we  obtain
 \begin{equation}
 \label{eq9}
 \frac{\ddot a}{a}=-\frac{8\pi L_p^{n-1}}{n(n-1)}\left[(n-2)\rho+np\right],
 \end{equation}
 With the continuity equation in $(n+1)$ dimensions, $\dot \rho +nH(\rho+p)=0$, integrating
 (\ref{eq9}) yields the standard Friedmann equation
 \begin{equation}
 \label{eq10}
 H^2 +\frac{k}{a^2}=\frac{16\pi L_p^{n-1}}{n(n-1)}\rho.
 \end{equation}
 It is direct to see that when $n=3$, our equations (\ref{eq9}) and (\ref{eq10})
 reduce to the ones (\ref{eq3}) and (\ref{eq4}), respectively. Here
 once again, $k$ is an integration constant.

 Note that (\ref{eq9}) and (\ref{eq10}) are dynamical equations of a FRW universe in the Einstein's
   general relativity. It is therefore quite interesting to see whether the above procedure works or
   not in the Gauss-Bonnet gravity and more general Lovelock gravity, the latter is a natural generalization of
   general relativity in higher dimensional spacetime. It is well-known that the entropy
   formula of black hole in the Gauss-Bonnet gravity no longer obeys the Bekenstein-Hawking area formula,
   instead it has an additional correction term~\cite{GB}
   \begin{equation}
   \label{eq11}
   S=\frac{A_+}{4L_p^{n-1}}\left(1+\frac{n-1}{n-3}\frac{2\tilde \alpha}{r_+^2}\right),
   \end{equation}
   where $A_+=n\Omega_n r_+^{n-1}$ is the horizon area of the black hole with horizon radius $r_+$ and
   ${\tilde \alpha}$ is the Gauss-Bonnet coefficient with dimension $[\rm length]^2$. Applying the
   entropy formula (\ref{eq11}) to the holographic surface in our setup, we assume that the effective area of the
   holographic surface is
   \begin{equation}
   \label{eq12}
   {\tilde A} =A\left (1+\frac{n-1}{n-3}\frac{2\tilde \alpha}{H^{-2}}\right),
   \end{equation}
   with $A=n\Omega_n/H^{n-1}$. Based on the relation between the volume $V$ and the area $A$ of an
   $n$-sphere with radius $R$, one has
   \begin{equation}
   \label{eq13}
   \frac{dV}{dA}=\frac{R}{n-1}.
   \end{equation}
   Then we further assume that the effective volume increase in the case of Gauss-Bonnet gravity
   satisfies
   \begin{eqnarray}
     \frac{d{\tilde V}}{dt} &=& \frac{1}{(n-1)H} \frac{d{\tilde A}}{dt},
     \label{eq14} \\
   &=&-\frac{n\Omega_n}{H^{n+1}}(1+2{\tilde\alpha}H^2)\dot {H},
   \label{eq15} \\
   &=&-\frac{n\Omega_n}{2H^{n+2}}(H^2+{\tilde \alpha}H^4)^{\cdot}.
   \label{eq16}
   \end{eqnarray}
   We suppose from (\ref{eq16}) that the number of degrees of freedom on the holographic surface
   is given by
   \begin{equation}
   \label{eq17}
   N_{\rm sur}=\alpha \frac{n\Omega_n}{H^{n+1}L_p^{n-1}}(H^2 +{\tilde \alpha}H^4),
   \end{equation}
   and in this case, the bulk degrees of freedom is still given by (\ref{eq8}). Putting (\ref{eq15}) and
   (\ref{eq17}) into (\ref{eq6}), we obtain
   \begin{equation}
   \label{eq18}
   (1+2{\tilde \alpha}H^2){\dot H} +(1+{\tilde \alpha}H^2)H^2 =-\frac{8\pi L_p^{n-1}}{n(n-1)}\left [(n-2)\rho +np\right ].
   \end{equation}
   Integrating (\ref{eq18}) with the continuity equation, we arrive
   at
   \begin{equation}
   H^2 +{\tilde \alpha}H^4 = \frac{16\pi L_p^{n-1}}{n(n-1)}\rho.
   \end{equation}
   Here we have set the integration constant to zero. This is
   nothing, but corresponding Friedmann equation of spatial flat FRW universe in the Gauss-Bonnet
   gravity~\cite{CaiKim}. Note that if one does not set the
   integration constant to zero, unlike the case in the Einstein's
   general relativity, it cannot be explained as the spatial
   curvature of the universe.

   In the more general Lovelcok gravity, the entropy of black hole has the following
   form~\cite{Cai3}
   \begin{equation}
   S= \frac{A_+}{4L_p^{n-1}}\sum^{m}_{i=1}\frac{i(n-1)}{(n-2i+1)}\hat{c_i}r_+^{2-2i},
   \end{equation}
   where $\hat c_i$ are some coefficients in front of Euler density terms in the
   theory and $m=[n/2]$. We suppose from the entropy expression
   that the effective area of the holographic surface is
   \begin{equation}
  \tilde {A}=\frac{n\Omega_n}{H^{n-1}}\sum^{m}_{i=1}\frac{i(n-1)}{(n-2i+1)}\hat{c_i}H^{2i-2},
   \end{equation}
   the increase of the effective volume is then given by
   \begin{eqnarray}
   \frac{d\tilde{V}}{dt}&=&\frac{1}{(n-1)H}\frac{d{\tilde A}}{dt},\\
   &=& -\frac{n\Omega_n}{H^{n+3}}\left (\sum^{m}_{i=1}i\hat{c_i}H^{2i}\right)\dot{H},
   \label{eq22}\\
   &=&-\frac{n\Omega_n}{2H^{n+2}}\left (\sum^m_{i=1}\hat{c_i}H^{2i}\right
   )^{\cdot}.
   \label{eq23}
   \end{eqnarray}
   In this case, we assume from (\ref{eq23}) that the number of the degrees of freedom on the holographic surface is
   \begin{equation}
   \label{eq24}
   N_{\rm sur}=\alpha \frac{n\Omega_n}{H^{n+1}L_p^{n-1}}\sum^m_{i=1}\hat{c_i}H^{2i}.
   \end{equation}
   Substituting (\ref{eq22}) and (\ref{eq24}) into (\ref{eq6}) together with the
   bulk degrees of freedom (\ref{eq8}), we arrive at
   \begin{equation}
   \left(\sum^m_{i=1}i \hat{c_i}H^{2i-2}\right)\dot {H}
   +\sum^m_{i=1}\hat{c_i}H^{2i}=-\frac{8\pi L_p^{n-1}}{n(n-1)}\left [(n-2)\rho+np\right].
   \end{equation}
   Integrating this equation with help of the continuity equation yields
   \begin{equation}
   \sum^m_{i=1}\hat{c_i}H^{2i}=\frac{16\pi L_p^{n-1}}{n(n-1)}\rho,
   \end{equation}
   where we have also set the integration constant to vanish. This
   equation is just  corresponding Friedmann equation of a flat FRW
   universe in the Lovelock gravity~\cite{CaiKim}. Note that here
   like the case of the Gauss-Bonnet gravity, the integration
   constant cannot be interpreted as the spatial curvature of the
   universe.

To summarize, in this short note we have investigated the idea
recently made by Padmanabhan~\cite{Pad12} that the emergence of
space and expansion of the universe are due to the difference
between the number of degrees of freedom on the holographic surface
and the one in the emerged bulk. In the setup of a higher
dimensional FRW universe, we have indeed derived the dynamical
equations of the universe.  From the entropy formulas of black hole
in the Gauss-Bonnet gravity and more general Lovelock gravity, by
properly modifying the volume increase  of the emerged space and the
number of degrees of freedom on the holographic surface, we have
also obtained correct Friedmann equations of a flat FRW universe in
those gravity theories. Finally  we would like to mention two pints
here. One is that in the case of Einstein's general relativity, from
the acceleration equation (\ref{eq9}) to the Friedmann equation
(\ref{eq10}), the integration constant $k$ can be interpreted as the
spatial curvature of the FRW universe, but it is no longer valid in
the cases of Gauss-Bonnet gravity and Lovelock gravity. The other is
that in the setting of this paper the expression of the number of
degrees of freedom on the holographic surface is not simply the same
as the one of black hole entropy, although they closely relate each
other, in the Gauss-Bonnet gravity and Lovelock gravity. These are
worthy further to investigate.

\section*{Acknowledgments}
The author would like to thank T. Padmanabhan for helpful
correspondences. This work was supported in part by the National
Natural Science Foundation of China (No.10821504, No.10975168 and
No.11035008), and in part by the Ministry of Science and Technology
of China under Grant No. 2010CB833004.



\begin{thebibliography}{99}
\bibitem{Sakharov}A.~D.~Sakharov,
  ``Vacuum quantum fluctuations in curved space and the theory of gravitation,''
  Sov.\ Phys.\ Dokl.\  {\bf 12}, 1040 (1968)
  [Dokl.\ Akad.\ Nauk Ser.\ Fiz.\  {\bf 177}, 70 (1967)]
  [Sov.\ Phys.\ Usp.\  {\bf 34}, 394 (1991)]
  [Gen.\ Rel.\ Grav.\  {\bf 32}, 365 (2000)].

\bibitem{Visser}M.~Visser,
  ``Sakharov's induced gravity: A Modern perspective,''
  Mod.\ Phys.\ Lett.\ A {\bf 17}, 977 (2002)
  [gr-qc/0204062].

\bibitem{Review}T.~Padmanabhan,
  ``Thermodynamical Aspects of Gravity: New insights,''
  Rept.\ Prog.\ Phys.\  {\bf 73}, 046901 (2010)
  [arXiv:0911.5004 [gr-qc]].

\bibitem{Jac}T.~Jacobson,
``Thermodynamics of space-time: The Einstein equation of state,''
Phys.\ Rev.\ Lett.\  {\bf 75}, 1260 (1995) [arXiv:gr-qc/9504004].

\bibitem{Pad09}T.~Padmanabhan,
  ``A Physical Interpretation of Gravitational Field Equations,''
  AIP Conf.\ Proc.\  {\bf 1241}, 93 (2010)
  [arXiv:0911.1403 [gr-qc]];
  T.~Padmanabhan,
  ``Entropy density of spacetime and thermodynamic interpretation of field equations of gravity in any diffeomorphism invariant theory,''
  arXiv:0903.1254 [hep-th].

\bibitem{Verlinde}E.~P.~Verlinde,
  ``On the Origin of Gravity and the Laws of Newton,''
  JHEP {\bf 1104}, 029 (2011)
  [arXiv:1001.0785 [hep-th]].

\bibitem{Unruh}W.~G.~Unruh,
  ``Notes on black hole evaporation,''
  Phys.\ Rev.\ D {\bf 14}, 870 (1976).

\bibitem{Pad10}T.~Padmanabhan,
  ``Equipartition of energy in the horizon degrees of freedom and the emergence of gravity,''
  Mod.\ Phys.\ Lett.\ A {\bf 25}, 1129 (2010)
  [arXiv:0912.3165 [gr-qc]].

\bibitem{Pad04}T.~Padmanabhan,
  ``Gravitational entropy of static space-times and microscopic density of states,''
  Class.\ Quant.\ Grav.\  {\bf 21}, 4485 (2004)
  [gr-qc/0308070].

\bibitem{CaiKim}R.~-G.~Cai and S.~P.~Kim,
  ``First law of thermodynamics and Friedmann equations of Friedmann-Robertson-Walker universe,''
  JHEP {\bf 0502}, 050 (2005)
  [hep-th/0501055].

\bibitem{CCH}R.~-G.~Cai, L.~-M.~Cao and Y.~-P.~Hu,
  ``Hawking Radiation of Apparent Horizon in a FRW Universe,''
  Class.\ Quant.\ Grav.\  {\bf 26}, 155018 (2009)
  [arXiv:0809.1554 [hep-th]].

\bibitem{CCN}R.~-G.~Cai, L.~-M.~Cao and N.~Ohta,
  ``Friedmann Equations from Entropic Force,''
  Phys.\ Rev.\ D {\bf 81}, 061501 (2010)
  [arXiv:1001.3470 [hep-th]].

\bibitem{Pad12}T.~Padmanabhan,
  ``Emergence and Expansion of Cosmic Space as due to the Quest for Holographic Equipartition,''
  arXiv:1206.4916 [hep-th].
\bibitem{GB}
R.~G.~Cai,
``Gauss-Bonnet black holes in AdS spaces,'' Phys.\ Rev.\
D {\bf 65}, 084014 (2002) [arXiv:hep-th/0109133];
R.~G.~Cai and Q.~Guo,
``Gauss-Bonnet black holes in dS spaces,''
Phys.\ Rev.\ D {\bf 69}, 104025 (2004) [arXiv:hep-th/0311020];


\bibitem{Cai3}
R.~G.~Cai,
``A note on thermodynamics of black holes in Lovelock
gravity,'' Phys.\ Lett.\ B {\bf 582}, 237 (2004)
[arXiv:hep-th/0311240].



\end{thebibliography}
\end{document}